# Stable ring vortex solitons in Bessel optical lattices

Yaroslav V. Kartashov,[1,2] Victor A. Vysloukh,[3] Lluis Torner[1]

[1]*ICFO-Institut de Ciencies Fotoniques, and Department of Signal Theory and Communications, Universitat Politecnica de Catalunya, 08034, Barcelona, Spain*

[2]*Physics Department, M. V. Lomonosov Moscow State University, 119899, Moscow, Russia*

[3]*Departamento de Fisica y Matematicas, Universidad de las Americas – Puebla, Santa Catarina Martir, 72820, Puebla, Mexico*

Stable ring vortex solitons, featuring a bright-shape, appear to be very rare in nature. However, here we show that they exist and can be made dynamically stable in defocusing cubic nonlinear media with an imprinted Bessel optical lattice. We find the families of vortex lattice solitons and reveal their salient properties, including the conditions required for their stability. We show that the higher the soliton topological charge, the deeper the lattice modulation necessary for stabilization.

Vortex solitons with a bright shape, i.e., screw topological phase dislocations embedded in a localized ring-shaped beam, might exist in different systems with focusing nonlinearities [1]. However, such solitons realize higher-order, excited states of the corresponding nonlinear systems; therefore they tend to be highly prone to azimuthal modulational instabilities that lead to their spontaneous self-destruction into vorticityless solitons [2]. This process has been observed experimentally in different settings [3], with the orbital angular momentum associated to the input vorticity being carried away by the soliton fragments that fly off the original ring. Homogeneous defocusing nonlinear media can support stable vortex solitons, but those have the form of dark-shaped beams, i.e., dislocations that seat on a background which extents to infinity along the transverse direction [4]. Localized, stable ring-vortex solitons in homogeneous media are only known to exist in models with competing cubic-quintic, or quadratic-cubic nonlinearities [5,6]. Nevertheless, such models are very challenging to implement in practice, as they would require very large light intensities where the higher-order nonlinearities are typically accompanied by additional dominant processes, like multi-photon absorption [7]. A successful alternative is the use of confined systems, such as graded-index optical fibers [8], trapped Bose-Einstein condensates [9], or nonlinear photonic crystals with defects [10], where stabilization of the ring-shaped beam is induced by the corresponding trapping potentials.

In this Letter we introduce a new approach to form stable nonlinear, ring-shaped vortices which is based on the concept of *Bessel optical lattices.* It is known that spatial modulation of the refractive index profoundly affects soliton properties [11-13]. It was demonstrated recently that optical lattices with flexibly controlled refractive index modulation depth and period can be induced optically in photorefractive materials [14-17]. To date, efforts have been devoted to the investigation of solitons supported by lattices with a rectangular or honeycomb symmetry that are induced by intersecting plane waves [14-18]. However, Bessel lattices with a radial shape are also possible [19]. Here we address Bessel lattices imprinted in defocusing cubic nonlinear media and uncover the properties of higher-order excited vortex soliton states supported by the structure. We show that vortex lattice solitons (VLS) exist in these lattices and that they can be made dynamically stable with suitable lattice strength. The VLS are the nonlinear continuation of the lattice modes, but in contrast to such modes, VLS might extent over several lattice rings thus featuring a multi-ring bright shape.

We consider light propagation along the $z$ axis in a bulk medium with the defocusing cubic nonlinearity and transverse modulation of refractive index described by nonlinear Schrödinger equation for the normalized complex field amplitude $q$:

$$i\frac{\partial q}{\partial \xi} = -\frac{1}{2}\left(\frac{\partial^2 q}{\partial \eta^2} + \frac{\partial^2 q}{\partial \zeta^2}\right) + q|q|^2 - pR(\eta,\zeta)q. \qquad (1)$$

The longitudinal $\xi$ and transverse $\eta,\zeta$ coordinates are scaled to the diffraction length and input beam width, respectively. The parameter $p$ is proportional to the depth of the refractive index modulation, and the function $R(\eta,\zeta) = J_1^2[(2b_{\text{lin}})^{1/2}r]$ with $r^2 = \eta^2 + \zeta^2$ stands for the transverse profile of refractive index; the parameter $b_{\text{lin}}$ is related to the radii of rings in the first-order Bessel lattice. The optical field of the lattice-crating first-order Bessel beam is given by $J_1[(2b_{\text{lin}})^{1/2}r]\exp(-ib_{\text{lin}}\xi + i\phi)$, where $\phi$ is the azimuthal angle. Such beams can be created experimentally by holographic techniques [20], while vectorial interactions in a slow Kerr-type medium (including, e.g. photorefractive crystals) can be utilized for trapping and guiding beams of orthogonal polarization in the lattice formed by the Bessel beam. We assume that the refractive index profile is given by intensity of the first-order Bessel beam as this case is favorable for vortex soliton formation (see Fig. 1(a)). The depth of refractive index modulation is assumed to be small, i.e., of the order of the nonlinear contribution. Eq. (1) requires nonlinearities of different signs for soliton and lattice-creating beams. In principle, such experimental conditions can be met in the photorefractive semiconductor crystals such as GaAs:Cr, InP:Fe, and CdTe:In, belonging to the $\overline{4}3m$ point symmetry group [21]. These materials are transparent for near infrared wavelengths, exhibit strong photorefractivity (e.g., $n^3r_{41} = 152$ pm/V in CdTe:In) and the sign of nonlinearity might be changed by a $\pi/2$-rotation of the polarization direction. Notice that the peak value of photorefractive contribution to the refractive index in such crystals could reach $\sim 10^{-3}$ (that corresponds to $p \sim 10$ in Eq. (1)) provided that sufficiently strong static electric field is applied. Eq. (1) admits several conserved quantities, including the energy flow $U = \int_{-\infty}^{\infty}\int_{-\infty}^{\infty}|q|^2 d\eta d\zeta$. We also stress that Eq. (1) holds for trapped suitable Bose-Einstein condensates with repulsive interactions [9].

We search for solutions of Eq. (1) in the form $q(\eta,\zeta,\xi) = w(r)\exp(im\phi)\exp(ib\xi)$, where $b$ is the propagation constant, $m$ is the topological charge, and $w(r)$ is the real function. Substitution of the light field in such form into Eq. (1) yields

$$\frac{d^2w}{dr^2} + \frac{1}{r}\frac{dw}{dr} - \frac{m^2w}{r^2} - 2bw - 2w^3 + 2pRw = 0 \qquad (2)$$

an equation that we solved numerically with a relaxation method. Mathematically, the soliton families are defined by the propagation constant $b$, the lattice depth $p$, and the parameter $b_{\text{lin}}$. Since one can use the scaling transformation $q(\eta,\zeta,\xi,p) \to \chi q(\chi\eta,\chi\zeta,\chi^2\xi,\chi^2p)$ to obtain various families of solitons from a given one, we set the transverse scale in such way that $b_{\text{lin}} = 2$ and varied $b$ and $p$. To analyze the dynamical stability of the soliton families we searched for perturbed solutions with the form $q(\eta,\zeta,\xi) = [w(r) + u(r,\xi)\exp(in\phi) + v^*(r,\xi)\exp(-in\phi)]\exp(ib\xi + im\phi)$, where the perturbation components $u,v$ could grow with complex rate $\delta$ upon propagation, and $n$ is the azimuthal index of the perturbation. Linearization of Eq. (1) around a stationary solution $w(r)$ yields the eigenvalue problem:

$$i\delta u = -\frac{1}{2}\left(\frac{d^2}{dr^2} + \frac{1}{r}\frac{d}{dr} - \frac{(m+n)^2}{r^2}\right)u + bu + w^2(v + 2u) - pRu,$$
$$-i\delta v = -\frac{1}{2}\left(\frac{d^2}{dr^2} + \frac{1}{r}\frac{d}{dr} - \frac{(m-n)^2}{r^2}\right)v + bv + w^2(u + 2v) - pRv, \quad (3)$$

which we solved numerically.

First we address the properties of the vorticityless solitons with zero topological charge $m = 0$ (Fig. 1), which physically correspond to the nonlinear continuation of the lowest-order mode confined by the lattice. The energy flow of such solitons is a monotonically decreasing function of the propagation constant (Fig. 1(b)). The energy flow goes to infinity at $b \to 0$ and vanishes at the upper cutoff $b_{\text{co}}$ of the propagation constant. Since the lattice profile has a local minimum at $r = 0$, vorticityless solitons have a small intensity dip on their top (Fig. 1(c)). At small power levels, when $b \to b_{\text{co}}$, vorticityless solitons transform into linear modes guided by the first lattice ring, while at $b \to 0$ where defocusing nonlinearity dominates the soliton diameter grows and it covers several lattice rings. As $b \to 0$ the soliton diameter increases dramatically while its maximal amplitude remains almost unchanged. The area of existence of vorticityless solitons broadens monotonically with growth of lattice depth (inset in Fig. 1(c)). Linear stability analysis revealed that vorticityless solitons solutions are stable in the entire domain of their existence, as expected on physical grounds. To confirm the results of the linear stability analysis we performed extensive numerical simulations using Eq. (1) with the perturbed input conditions $q|_{\xi=0} = w(r)[1 + \rho(r,\phi)]$, where $w(r)$ describes stationary soliton, and $\rho(r,\phi)$ is the random function with Gaussian distribution and variance $\sigma_{\text{noise}}^2$. An example of stable propagation of a perturbed soliton is shown in Fig. 1(d).

We now consider VLS with topological charge $m=1$ (Fig. 2). Note that such lattice solitons, as well as the vorticityless ones, exist because defocusing nonlinearity and diffraction are balanced by the lattice that focuses radiation into the region with higher refractive index. Therefore, the Bessel lattice affords confinement of light that is impossible in a uniform defocusing medium. The energy flow of the VLS is a monotonically decreasing function of the propagation constant (Fig. 2(a)). There exist zero lower and positive upper cutoffs on the propagation constant. The energy flow of the VLS diverges as $b \to 0$ and vanishes as $b$ approaches the upper cutoff $b_{co}$. With increase of the energy flow, the VLS get wider and cover many lattice rings (Fig. 2(b)). The existence domain of VLS with unit topological charge is displayed in Fig. 2(c). The width of the existence domain increases monotonically with growth of the lattice depth. Notice that at fixed lattice depth $p$, the width of the existence domain on propagation constant reaches its maximal value for vorticityless solitons and decreases with growth of the soliton topological charge.

The central result of this Letter is that the VLS become dynamically stable in suitable domains of their existence. This is depicted in Fig. 2(c). In the plot we show the critical value of propagation constant $b_{cr}^n$ above which no perturbations with the azimuthal index $n$ and nonzero real part of growth rate were found. Notice that the precise structure of instability regions (shaded area in Fig. 2(c)) is quite complicated. There exist multiple narrow stability "windows" near the upper cutoff even for shallow lattices, but we do not show them here because stabilization close to the upper cutoff (where soliton transforms into the linear mode guided by first lattice ring) is not surprising. Our simulations indicate that VLS with $m=1$ from the shaded area in Fig. 2(c) self-destroy under the action of perturbation with the azimuthal index $n=1$, while the corresponding instability is of a oscillatory type with $\mathrm{Re}(\delta) \ll \mathrm{Im}(\delta)$. Typically decay of the unstable vortices in Bessel lattice occurs at hundreds of propagation units. Decay of the unstable VLS either produces diffracting radiation or sets of vorticityless solitons. We found that when the lattice depth exceeds a critical value, of about $p_{cr} \approx 12.9$, the instability regions *cease to exist* and VLS become stable in the *entire domain* of their existence. The *stable* propagation of a vortex lattice soliton with $m=1$ in the presence of broadband input noise is illustrated in Fig. 2(d). To illustrate that the defocusing nonlinearity is a necessary ingredient for the existence of the stable VLS, Fig.

3 shows the three-dimensional shapes of different VLS together with the linear mode supported by the lattice. Notice the differences in beam profile introduced by the defocusing nonlinearity, in particular the multi-ring structure acquired by the VLS at moderate and high energy flows.

Vortex lattice solitons with topological charge $m=2$ were also studied. Their properties have much in common with properties of solitons with $m=1$ and are summarized in Fig. 4. The instability domain is located near lower cutoff on propagation constant. It also has complex structure with separate stability "windows" (not shown at the plot), and its width decreases monotonically with growth of the lattice depth. Notice that vortices from shaded area in Fig. 4(c) are affected by perturbations with azimuthal indexes $n=1,2$, which indicates that spectrum of harmful perturbations enriches with growth of the vortex topological charge. However, the important result uncovered is that for deep enough lattices a broad stability domain appears too, as shown in Fig. 4(c). An illustrative example of the propagation of a stable ring-shaped vortex soliton with $m=2$ is presented in Fig. 4(d). We considered also vortex solitons with higher topological charges and found stability conditions. Consistent with intuitive expectations, a general conclusion can be drawn that the higher is the topological charge of the vortex the deeper is the lattice required for its stabilization. It is worth mentioning that we also found higher-order "twisted" vortex and vorticityless modes whose field $w(r)$ alternates on successive lattice rings, but their stability analysis is beyond the scope of this Letter. Finally, we also studied VLS in Bessel lattices imprinted in focusing nonlinear media, but we found them all to be unstable against azimuthal modulational instabilities.

To summarize, we stress that optical lattices induced by nondiffracting first-order Bessel beam in media with defocusing nonlinearity have been shown to be able to support stable vorticityless and also stable multi-ring vortex solitons. The predicted existence of the bright-shaped vortex lattice solitons, which do not self-destroy by azimuthal modulational instabilities, is a rare phenomenon in physics and it is an important example of the unique phenomena afforded by Bessel optical lattices. The results reported are relevant not only for nonlinear optics but also for suitable Bose-Einstein condensates trapped in Bessel lattices with repulsive interatomic interactions.

This work has been partially supported by the Generalitat de Catalunya and by the Spanish Government through grant BFM2002-2861.

# Figure captions

Figure 1. (a) Bessel lattice of the first order. Regions with higher refractive index are shown with white color and regions with lower refractive index are shown with black color. (b) Energy flow of vorticityless soliton versus propagation constant for different lattice depths. (c) Amplitude profiles of solitons corresponding to points marked by circles in (b). Inset in (c) shows upper cutoff on propagation constant versus lattice depth. (d) Stable propagation of the vorticityless soliton with $b=0.8$ at $p=15$ in the presence of white noise with variance $\sigma_{\text{noise}}^2 = 0.01$. Cut of intensity distribution at $\zeta=0$ is shown.

Figure 2. (a) Energy flow of vortex solitons with $m=1$ versus propagation constant for different lattice depths. (b) Amplitude profiles of vortex solitons corresponding to points marked by circles in (a). (c) Areas of stability and instability (shaded) on the $(p,b)$ plane for vortex solitons with $m=1$. (d) Stable propagation of vortex with $m=1$, $b=0.7$ at $p=15$ in the presence of white noise with variance $\sigma_{\text{noise}}^2 = 0.01$. Cut of vortex intensity distribution at $\zeta=0$ is shown.

Figure 3. Three-dimensional intensity distributions for different VLS with $m=1$. (a) Energy flow $U=9$, lattice depth $p=15$. (b) $U=62$, $p=15$. (c) $U=124$, $p=15$. (d) $U=62$, $p=10$. The VLS shown in (a) corresponds to the quasi-linear regime. All figures plotted with the same scale in the vertical axis.

Figure 4. (a) Energy flow of vortex solitons with $m=2$ versus propagation constant for different lattice depths. (b) Amplitude profiles of vortex solitons corresponding to points marked by circles in (a). (c) Areas of stability and instability (shaded) on the $(p,b)$ plane for vortex solitons with $m=2$. (d) Stable propagation of vortex with $m=2$, $b=1.6$ at $p=40$ in the presence of white noise with variance $\sigma^2_{\text{noise}} = 0.01$. Cut of vortex intensity distribution at $\zeta = 0$ is shown.

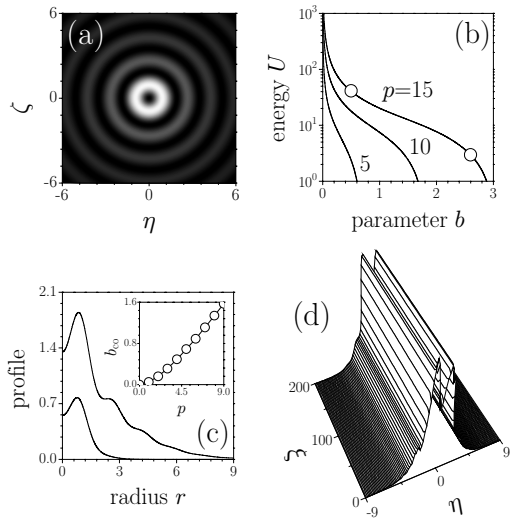

Figure 1. (a) Bessel lattice of the first order. Regions with higher refractive index are shown with white color and regions with lower refractive index are shown with black color. (b) Energy flow of vorticityless soliton versus propagation constant for different lattice depths. (c) Amplitude profiles of solitons corresponding to points marked by circles in (b). Inset in (c) shows upper cutoff on propagation constant versus lattice depth. (d) Stable propagation of the vorticityless soliton with $b = 0.8$ at $p = 15$ in the presence of white noise with variance $\sigma_{\text{noise}}^2 = 0.01$. Cut of intensity distribution at $\zeta = 0$ is shown.

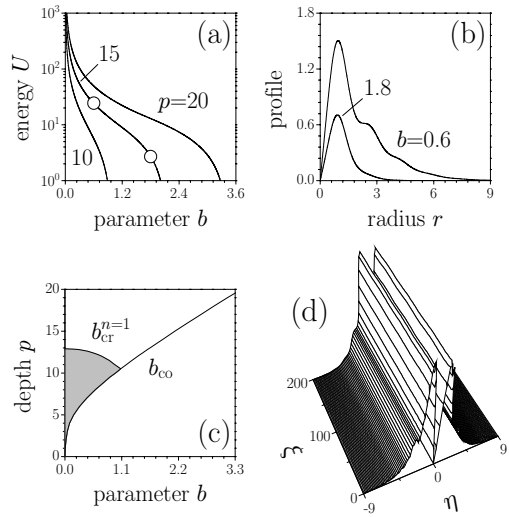

Figure 2.  (a) Energy flow of vortex solitons with $m=1$ versus propagation constant for different lattice depths. (b) Amplitude profiles of vortex solitons corresponding to points marked by circles in (a). (c) Areas of stability and instability (shaded) on the $(p,b)$ plane for vortex solitons with $m=1$. (d) Stable propagation of vortex with $m=1$, $b=0.7$ at $p=15$ in the presence of white noise with variance $\sigma_{\text{noise}}^2 = 0.01$. Cut of vortex intensity distribution at $\zeta = 0$ is shown.

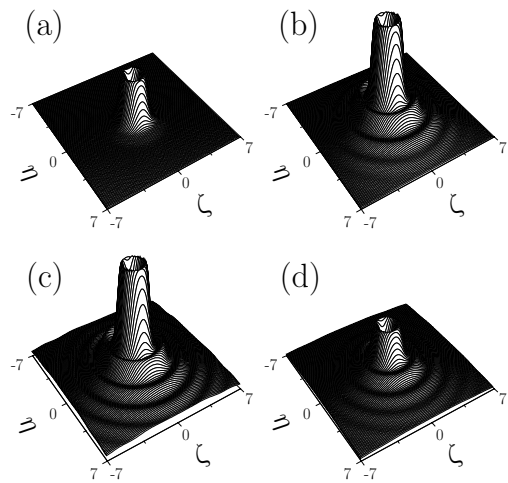

Figure 3. Three-dimensional intensity distributions for different VLS with $m=1$. (a) Energy flow $U=9$, lattice depth $p=15$. (b) $U=62$, $p=15$. (c) $U=124$, $p=15$. (d) $U=62$, $p=10$. The VLS shown in (a) corresponds to the quasi-linear regime. All figures plotted with the same scale in the vertical axis.

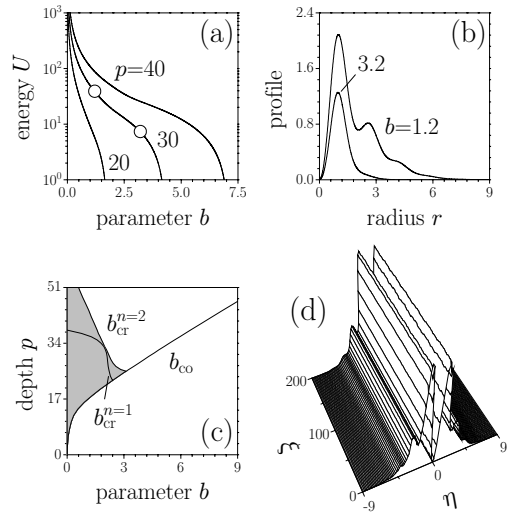

Figure 4. (a) Energy flow of vortex solitons with $m=2$ versus propagation constant for different lattice depths. (b) Amplitude profiles of vortex solitons corresponding to points marked by circles in (a). (c) Areas of stability and instability (shaded) on the $(p,b)$ plane for vortex solitons with $m=2$. (d) Stable propagation of vortex with $m=2$, $b=1.6$ at $p=40$ in the presence of white noise with variance $\sigma_{\text{noise}}^2 = 0.01$. Cut of vortex intensity distribution at $\zeta = 0$ is shown.